\documentclass[pre,twocolumn,showpacs,floatfix]{revtex4}
\usepackage{graphicx}
\begin{document}

\title{Simulational nanoengineering: Molecular dynamics implementation of an
atomistic Stirling engine}

\author{D. C. Rapaport}
\email{rapaport@mail.biu.ac.il}
\affiliation{Physics Department, Bar-Ilan University, Ramat-Gan 52900, Israel}

%\date{\today}
%\date{January 20, 2009}
%\date{March 02, 2009}
\date{April 05, 2009}

\begin{abstract}
A nanoscale-sized Stirling engine with an atomistic working fluid has been
modeled using molecular dynamics simulation. The design includes heat exchangers
based on thermostats, pistons attached to a flywheel under load, and a
regenerator. Key aspects of the behavior, including the time-dependent flows,
are described. The model is shown to be capable of stable operation while
producing net work at a moderate level of efficiency.
\end{abstract}

\maketitle

The Stirling engine, an external combustion engine invented almost two centuries
ago, and an early competitor of the steam engine, continues to attract interest
owing to its potential effectiveness as a power source and its success in
specialized applications \cite{uri84,wes86,zem68}. The underlying thermodynamic
cycle consists of four stages: low-temperature isothermal compression of the
working fluid, constant-volume displacement of the fluid between the cold and
hot spaces of the engine with (optionally) the fluid gaining heat while passing
through a heat-storing regenerator, high-temperature isothermal expansion, and
constant-volume displacement between the hot and cold spaces with heat being
returned to the regenerator. The net work is the excess energy produced by the
expanding hot fluid over that needed to recompress the cold fluid. Unlike
internal combustion engines, there is complete flexibility as to fuel type,
including even solar energy; the fact that the complexities of combustion
\cite{man08} are avoided makes the Stirling engine an attractive candidate for
simulation.

This paper describes the molecular dynamics (MD) simulation
of a simplified Stirling engine using a discrete-particle working fluid. MD
\cite{ald57,rah64,rap04}
provides the capability for direct atomistic modeling of nanomachinery, together
with the accompanying complex thermodynamic and fluid-dynamic processes. While
past MD studies have included instances of heat production as a byproduct of
work, e.g., in fracture \cite{hol95}, Taylor--Couette flow \cite{hir98} and
fluid jets \cite{mos00}, the present simulations extend the scope of MD to
systems that harness thermal energy for producing useful
work\footnote{Thermodynamics was developed to explain the engines of the
industrial revolution, statistical mechanics provides the atomistic basis for
equilibrium thermodynamics, and MD addresses nonequilibrium phenomena beyond the
scope of statistical mechanics; the present study revisits the original problem
with a mechanical perspective.}.

The model engine incorporates a number of features: pistons driven by collisions
with the fluid atoms, heat input and output controlled by thermostats that
maintain isothermal conditions in the hot and cold spaces, a rotating flywheel
connected to the pistons and subject to an applied load, and an atomistic
regenerator intended to reduce heat wastage. The working fluid experiences
substantial temperature and pressure variation over the cycle, but its
thermodynamic and heat/mass transport properties are intrinsic to the atomistic
model. This represents an advantage of MD over the alternative continuum-based
analysis \cite{mar83} where such details must be provided separately; if these
and subsequent more extensive studies can be quantitatively validated, the role
of MD in nanoengineering is assured.

\begin{figure}
\includegraphics[scale=0.35]{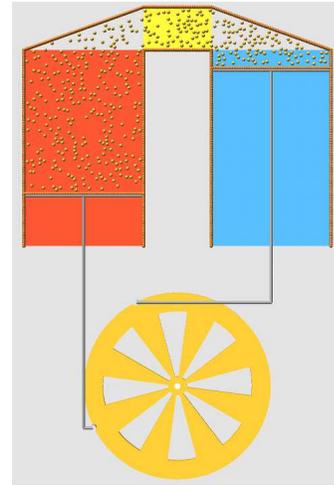}
\caption{\label{fig:1} (Color online) Model Stirling engine, showing hot
(left side) and cold spaces, regenerator (top center), pistons, counterclockwise
rotating flywheel, linkages, and working fluid.}
\end{figure}

There are different Stirling engine designs, each involving engineering
considerations that are not addressed here. It is the two-piston {\em alpha}
version, shown in Fig.~\ref{fig:1}, that is modeled (the {\em beta} version,
with a single piston and a mechanical displacer, is one of the alternatives).
Even though the simulations consider a system that is an idealization of a real
Stirling engine, its relative complexity -- by MD standards -- calls for several
implementation decisions, with the goal of achieving a simple but reasonably
complete operational model. The two-dimensional case is considered for
convenience.

The atoms of the working fluid are soft-disk particles, interacting through a
short-range potential $U(r) = 4\epsilon [(\sigma/r)^{12}-(\sigma/r)^6]+\epsilon$
with $r<r_c=2^{1/6} \sigma$ (condensation is avoided since there is no
attraction); similar disks line the fixed engine walls and the moving piston
faces. The linkages connecting the pistons to the flywheel ensure sinusoidal
piston motion, but are idealized since the transverse parts are of variable
length; the phase angle between the pistons is fixed at $90^\circ$. The fluid
atoms and movable engine components obey Newtonian dynamics; the force
evaluation and solution of the equations of motion employ standard MD techniques
\cite{rap04}.

In reduced MD units, the model engine has stroke $2 R_f=112.4$, where $R_f$ is
the flywheel radius and bore 67.7 (in MD units where, typically, $\sigma=1$
corresponds to a length unit of 3.4{\AA}, $\epsilon=1$ to a time unit of 2.16
psec if fluid atoms have mass $m=1$, and $k_B=1$ defines temperature); the
actual value of $R_f$ amounts to a mere 191\AA. The working fluid contains
$N_a=1368$ atoms; this `large' system is the main focus of the analysis, but a
half-size `small' system (shown in Fig.~\ref{fig:1}) with $N_a=407$ is also
considered. The minimal size enables runs of adequate length without excessive
computation, while demonstrating the extreme scales at which behavior remains
reasonable.

Heat exchangers responsible for thermal input and output are modeled using
thermostats. These act separately over the hot and cold spaces swept by the
pistons, are applied every time step, and entail rescaling the velocities of all
the fluid atoms in each space, after allowing for the current average flow. Use
of thermostats, rather than thermalizing walls, prevents temperature
fluctuations and enhances heat transfer (since unrestricted transfer rates from
and to exogenous heat reservoirs are implied).

The flywheel phase angle $\theta$ is defined so that the hot piston is at its
top position when $\theta=0^\circ$. Average fluid density $N_a/A(\theta)$, where
$A(\theta)$ is the area occupied by the fluid, varies between 0.256 and 0.085,
corresponding to a 3.03:1 compression ratio. The inertial contribution of the
drive mechanism is assigned to the flywheel, with pistons and linkages assumed
massless. The flywheel moment of inertia $I_f = (\pi/2) \rho_f R_f^4$; for
density $\rho_f = 10^4$ the rotation speed $\omega$ is low enough that piston
speeds ($\le R_f \omega$) are typically an order of magnitude less than the mean
thermal velocity of the fluid atoms $v_t$ (the resulting slow fluid flow
justifies the thermostat implementation). The flywheel kinetic energy also
exceeds that of the fluid by a similar factor, ensuring minimal fluctuations in
$\omega$.

The role of the regenerator was recognized by its inclusion in the early
Stirling design. Although not essential for operation, its function is to
improve heat utilization by extracting heat from the hot fluid as it flows to
the cold space and returning it when flow is reversed; typically implemented as
a metallic mesh, it also impedes the flow. The model regenerator is assembled
from $N_r=32$ atoms with $m=50$, tethered to a uniform grid by a potential $50
U(r_c-r)$, where $r$ is the distance from the relevant grid site. Transiting
fluid atoms collide with the (unthermostatted) tethered atoms and heat is
transferred to or from the regenerator, depending on the relative temperatures;
the heavier atoms undergo limited, relatively slow motion, but the contribution
to performance in terms of heat storage is minimal since, to lessen flow
impedance, $N_r \ll N_a$. The area occupied by the regenerator is connected to
the hot and cold spaces by triangular ducts. 

Work is performed against an applied torque load $\gamma R_f \omega$, with
$\gamma=100$. There is no energy loss due to the rough boundary walls since all
collisions are elastic (energy is conserved when thermostats and load are
removed). Temperature in the hot space $T_h$ is a parameter, and the cold space
$T_c$ is fixed at unity. The system is started at maximum compression with the
pistons at equal height ($\theta=45^\circ$) and rotation speed $\omega_0$. Run
length is $2 \times 10^9$ time steps (of size 0.005), so while all runs cover
the same time period, the number of cycles differs.

\begin{figure}
\includegraphics[scale=0.66]{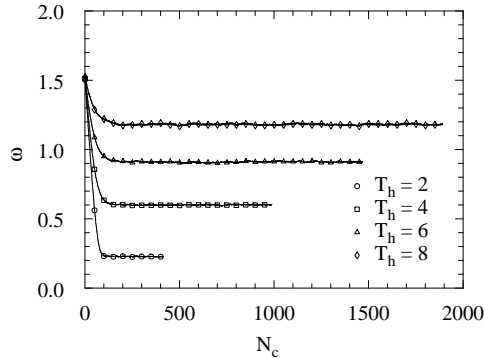}
\caption{\label{fig:2} Average rotation speed $\bar\omega$ (mrad/time), for runs
at different temperatures $T_h$; $N_c$ is the number of cycles.}
\end{figure}

\begin{figure}
\includegraphics[scale=0.66]{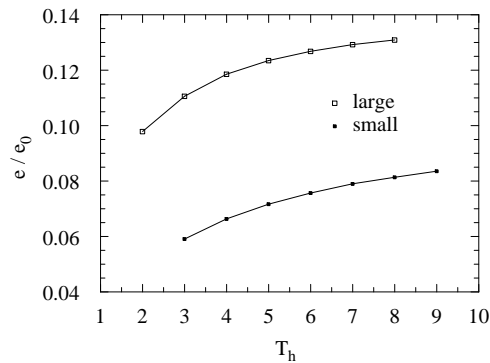}
\caption{\label{fig:3} Dependence of relative efficiency $e/e_0$ on $T_h$ for
large and small systems.}
\end{figure}

The principal outcome of the present simulations is a demonstration of the
operational capability of the engine. Fig.~\ref{fig:2} shows the cycle-averaged
rotation speed $\bar\omega$ as a function of the number of cycles $N_c$, for
runs at different $T_h$ (in real units, typical speeds are $\simeq 5 \times
10^7$ rot/s). $\bar\omega$ converges to a steady value following a transient
stage lasting $<200$ cycles, with no long-term drift and only small
size-dependent fluctuations. In addition to $\bar\omega$ increasing with $T_h$,
it also depends on the other parameters: for example, a higher load ($\gamma$),
but not too high to prevent rotation, reduces $\bar\omega$; the same is true if
regenerator collisional flow impedance is increased by raising $N_r$;
$\bar\omega$ increases with fluid density (by varying $N_a$) over a reasonable
range. The choice of $\omega_0$ (even $\omega_0=0$) does not affect the outcome,
provided the flywheel is able to complete the initial rotations, otherwise there
are damped oscillations about the state of maximum expansion. The overall
behavior of the small system is similar, though with enhanced fluctuations.

\begin{figure}
\includegraphics[scale=0.66]{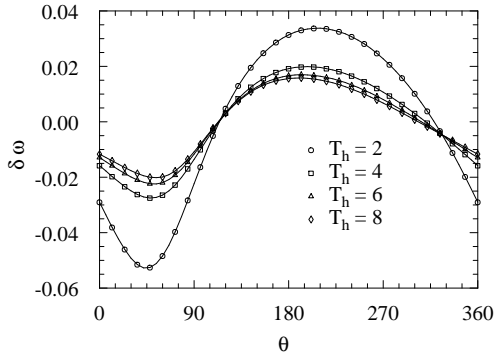}
\caption{\label{fig:4} Deviation from average rotation speed $\delta \omega$ as
a function of phase angle $\theta$.}
\end{figure}

\begin{figure}
\includegraphics[scale=0.66]{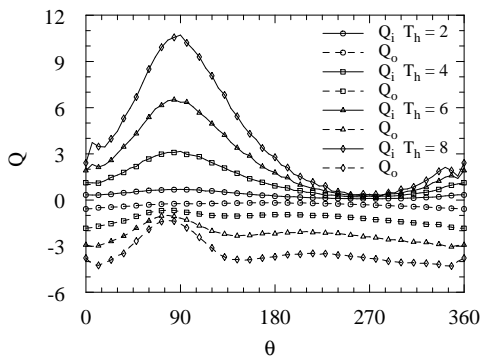}
\caption{\label{fig:5} Heat input and output, $Q_i(\theta)$ and $-Q_o(\theta)$
(the latter negative for clarity), over the cycle.}
\end{figure}

\begin{figure}
\includegraphics[scale=0.66]{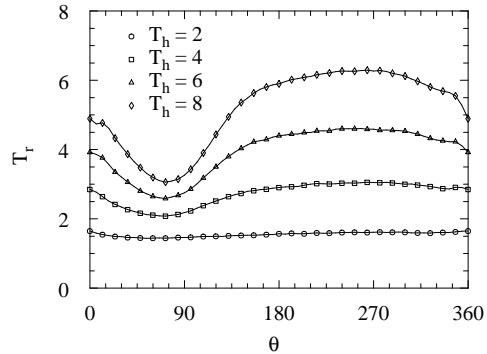}
\caption{\label{fig:6} Regenerator temperature $T_r(\theta)$.}
\end{figure}

\begin{figure}
\includegraphics[scale=0.66]{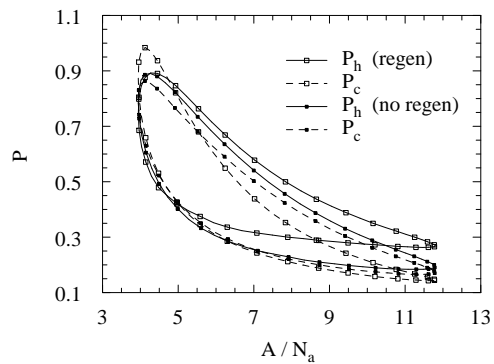}
\caption{\label{fig:7} Pressure $P(\theta)$ as a function of fluid area
$A(\theta)/N_a$, with and without regenerator ($T_h=4$).}
\end{figure}

Engine performance is expressed in terms of thermodynamic efficiency, the ratio
of net work $W$ to heat input $Q_i$; the latter represents the true operational
cost, whereas heat output to the environment $Q_o$ is waste. $Q_i$ and $Q_o$ are
the measured kinetic energy changes produced by the thermostats, and the source
of $W$ is the resistive torque on the flywheel; since there are no other loss
mechanisms, $\bar{W} = \bar{Q_i}-\bar{Q_o}$, where the averaging (over the
entire run, excluding the initial transient stage) allows for fluctuations.
Efficiency $e$ is expressed relative to the thermodynamic (Carnot) limiting
value $e_0 = 1-T_c/T_h$. Fig.~\ref{fig:3} shows the relative efficiency $e/e_0$
as a function of $T_h$, for both large and small systems
(size dependence is significant at such scales),
with values reaching a
respectable 0.13 (13\%). The presence of the regenerator degrades performance,
and when removed $e/e_0$ is increased to 0.24 at $T_h=4$; higher $\gamma$ also
raises $e$ since more work is done.

The dependence of rotation speed (run-averaged) on $\theta$ is shown in
Fig.~\ref{fig:4}; since the variation is weak (greater, even twofold, variation
occurs in the small system), the deviation $\delta\omega(\theta) =
\omega(\theta)- \bar\omega$ is shown. Speed is nonsinusoidal, slowest at $\theta
\simeq 45^\circ$ at maximum overall compression, and fastest at $\theta \simeq
200^\circ$ approaching maximum expansion; during the power portion of the cycle,
$d\omega/d\theta>0$ when work performed by the engine exceeds load. The range of
$\delta\omega$ is reduced at higher $T_h$; likewise if $I_f$ is increased.

\begin{figure*}
\includegraphics[scale=0.13]{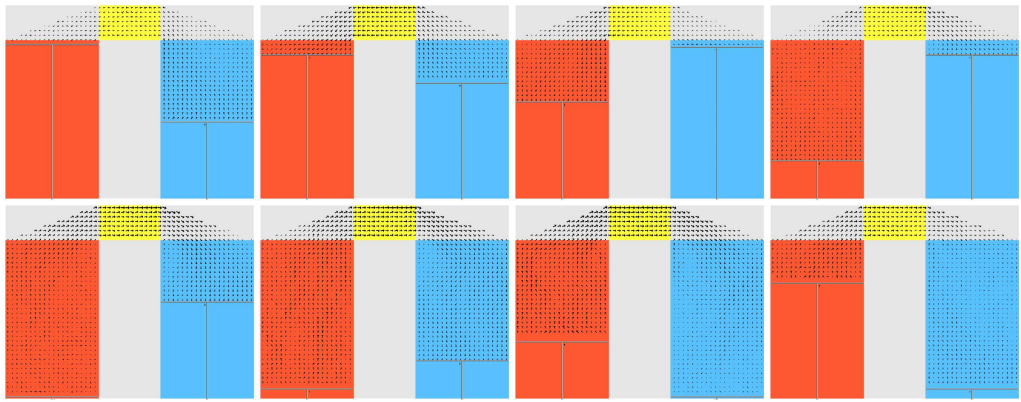}
\caption{\label{fig:8} (Color online) Fluid flow during the cycle (from upper
left, $\theta=0^\circ$, $45^\circ$,\ldots); to resolve the wide range of flow
speeds, arrow size $\propto \mathrm{(speed)}^{1/2}$ (slower flows appear
faster). Flow features are discussed in the text.}
\end{figure*}

Further measurements of $\theta$ dependence provide a more detailed picture of
the behavior. Fig.~\ref{fig:5} shows heat input and output through the
thermostats, $Q_i(\theta)$ and $Q_o(\theta)$; due to significant fluctuations,
averaging over multiple cycles is again necessary. The details depend on $T_h$,
with maximal $Q_i(\theta)$ at $\theta \simeq 90^\circ$, close to where
$|Q_o(\theta)|$ is at a minimum, and a value near zero at $\theta \simeq
270^\circ$; both also exhibit small irregularities in the vicinity of $\theta
\simeq 0^\circ$ where the fluid is compressed and the hot space practically
empty.

The effective regenerator temperature $T_r(\theta)$ is derived from the average
kinetic energy of the tethered atoms; the variation over the cycle is shown in
Fig.~\ref{fig:6}. For $T_h>2$ there is a minimum at $\theta \simeq 75^\circ$ as
the cold piston approaches the top position, and a broader maximum around
$\theta \simeq 270^\circ$; both limiting values increase with $T_h$. Small
irregularities appear near $0^\circ$, as before, that are consequences of the
model design and more pronounced in the small system. The temperature variation
shows that the regenerator responds correctly, although the capacity for heat
storage is very limited. Heat throughput is increased (together with $e$) when
the regenerator is removed to lower flow impedance.

Fluid pressure is evaluated from the virial; Fig.~\ref{fig:7} shows pressure
variation over the cycle in the hot and cold spaces, $P_h(\theta)$ and
$P_c(\theta)$, as a function of fluid area $A(\theta)/N_a$, for runs at $T_h=4$
with and without the regenerator; although similar to an indicator diagram
\cite{zem68}, the area enclosed does not correspond to $W$ away from
equilibrium. The general absence of pressure equalization between hot and cold
spaces is primarily due to regenerator flow impedance, with $P_c^{max}/P_c^{min}
\simeq 2 P_h^{max}/P_h^{min}$; without the regenerator $P_h(\theta) \simeq
P_c(\theta)$. Pressure dependence on $\theta$ (not shown) is not sinusoidal,
with a range of variation that increases with $T_h$; maxima are at $\theta$
between $50^\circ$ and $80^\circ$, in advance of the $\omega(\theta)$ inflection
point at $\simeq 120^\circ$ where power output begins to decline, with
$P_c^{max}$ preceding $P_h^{max}$ by $\simeq 10^\circ$.

Finally, at the most detailed level of examination, the cyclically changing
fluid flow is evaluated from spatially coarse-grained velocity fields
accumulated for $\theta$ segments of range $15^\circ$ and averaged over 100
successive cycles to reduce noise (some small variation persists even after
averaging). The frames of Fig.~\ref{fig:8} show every third segment (at
$T_h=4$). Since bulk flow rates are smaller than $v_t$ by at least an order of
magnitude, certain flow features only become apparent after averaging: because
of thermal expansion, at $135^\circ$ (frame 4), fluid can be seen leaving the
hot space and flowing back through the regenerator, even though the hot piston
is still descending, with the reverse occurring at $315^\circ$ (frame 8); there
is also a vortex that forms near the top of the inner wall of the cold space
while the cold piston is descending (frames 5-7).

\bibliography{stirleng}

\end{document}